\documentclass[aps,prd,amsfonts,amsmath,nofootinbib,tightenlines,preprint,showpacs,longbibliography,superscriptaddress]{revtex4-2}

\usepackage{graphicx}
\usepackage{xcolor}
\usepackage{hyperref}

\def\be{\begin{equation}}
\def\ee{\end{equation}}

\def\yr{\text{yr}}
\def\CGW{C^{\text{GW}}}
\def\CpGW{C'^{\text{GW}}}
\def\Cred{C^{\text{red}}}
\def\Cpred{C'^{\text{red}}}
\def\KGW{K^{\text{GW}}}
\def\Kred{K^{\text{red}}}
\def\phiGW{{\phi_{\text{GW}}}}
\def\phired{{\phi_{\text{red}}}}
\def\Nfreq{{N_{\text{freq}}}}
\def\NTOA{{N_{\text{TOA}}}}

\def\tGW{t_{\text{GW}}}
\def\tdet{t_{\text{det}}}
\def\tred{t_{\text{red}}}
\def\twhite{t_{\text{white}}}
\def\cZ{\mathcal{Z}}

\makeatletter
\let\ftype@table\ftype@figure
\makeatother

\begin{document}

\title{Evaluating the Fourier Approximation in Pulsar Timing Array Analysis}

% I edited this file by hand to remove ORCids because PRD did not want them.
% I also updated affiliations for Tim Dolch
\author{Yongqi Zhang}
\affiliation{Institute of Cosmology, Department of Physics and Astronomy, Tufts University, Medford, MA 02155, USA}
\affiliation{Cornell Center for Astrophysics and Planetary Science and Department of Astronomy, Cornell University, Ithaca, NY 14853, USA}
%\email{yongqi.zhang@nanograv.org}
\author{Hayden Scholz}
\affiliation{Institute of Cosmology, Department of Physics and Astronomy, Tufts University, Medford, MA 02155, USA}
%\email{hayden.scholz@nanograv.org}
\author{Ken D. Olum}
\affiliation{Institute of Cosmology, Department of Physics and Astronomy, Tufts University, Medford, MA 02155, USA}
%\email{ken.olum@nanograv.org}
\author{Lucas Steinberger}
\affiliation{Institute of Cosmology, Department of Physics and Astronomy, Tufts University, Medford, MA 02155, USA}
%\email{lucas.steinberger@nanograv.org}
\author{Gabriella Agazie}
\affiliation{Center for Gravitation, Cosmology and Astrophysics, Department of Physics and Astronomy, University of Wisconsin-Milwaukee,\\ P.O. Box 413, Milwaukee, WI 53201, USA}
%\email{gabriella.agazie@nanograv.org}
\author{Akash Anumarlapudi}
\affiliation{Department of Physics and Astronomy, University of North Carolina, Chapel Hill, NC 27599, USA}
%\email{akasha@unc.edu}
\author{Anne M. Archibald}
\affiliation{Newcastle University, NE1 7RU, UK}
%\email{anne.archibald@nanograv.org}
\author{Zaven Arzoumanian}
\affiliation{X-Ray Astrophysics Laboratory, NASA Goddard Space Flight Center, Code 662, Greenbelt, MD 20771, USA}
%\email{zaven.arzoumanian@nanograv.org}
\author{Paul T. Baker}
\affiliation{Department of Physics and Astronomy, Widener University, One University Place, Chester, PA 19013, USA}
%\email{paul.baker@nanograv.org}
\author{Paul R. Brook}
\affiliation{Institute for Gravitational Wave Astronomy and School of Physics and Astronomy, University of Birmingham, Edgbaston, Birmingham B15 2TT, UK}
%\email{paul.brook@nanograv.org}
\author{H. Thankful Cromartie}
\affiliation{National Research Council Research Associate, National Academy of Sciences, Washington, DC 20001, USA resident at Naval Research Laboratory, Washington, DC 20375, USA}
%\email{thankful.cromartie@nanograv.org}
\author{Kathryn Crowter}
\affiliation{Department of Physics and Astronomy, University of British Columbia, 6224 Agricultural Road, Vancouver, BC V6T 1Z1, Canada}
%\email{kathryn.crowter@nanograv.org}
\author{Megan E. DeCesar}
\affiliation{Department of Physics and Astronomy, George Mason University, Fairfax, VA 22030, resident at the U.S. Naval Research Laboratory, Washington, DC 20375, USA}
%\email{megan.decesar@nanograv.org}
\author{Paul B. Demorest}
\affiliation{National Radio Astronomy Observatory, 1003 Lopezville Rd., Socorro, NM 87801, USA}
%\email{paul.demorest@nanograv.org}
\author{Timothy Dolch}
\affiliation{Department of Physics and Astronomy, University of New
Mexico, Albuquerque, NM 87131, USA}
\affiliation{Department of Physics, Hillsdale College, 33 E. College
Street, Hillsdale, MI 49242, USA}
\affiliation{Eureka Scientific, 2452 Delmer Street, Suite 100, Oakland,
CA 94602-3017, USA}
\affiliation{SETI Institute, 339 N Bernardo Ave Suite 200, Mountain
View, CA 94043, USA}
%\email{timothy.dolch@nanograv.org}
\author{Justin A. Ellis}
\altaffiliation{Infinia ML, 202 Rigsbee Avenue, Durham NC, 27701, USA}
%\email{justin.ellis@nanograv.org}
\author{Elizabeth C. Ferrara}
\affiliation{Department of Astronomy, University of Maryland, College Park, MD 20742, USA}
\affiliation{Center for Research and Exploration in Space Science and Technology, NASA/GSFC, Greenbelt, MD 20771}
\affiliation{NASA Goddard Space Flight Center, Greenbelt, MD 20771, USA}
%\email{elizabeth.ferrara@nanograv.org}
\author{William Fiore}
\affiliation{Department of Physics and Astronomy, University of British Columbia, 6224 Agricultural Road, Vancouver, BC V6T 1Z1, Canada}
%\email{william.fiore@nanograv.org}
\author{Emmanuel Fonseca}
\affiliation{Department of Physics and Astronomy, West Virginia University, P.O. Box 6315, Morgantown, WV 26506, USA}
\affiliation{Center for Gravitational Waves and Cosmology, West Virginia University, Chestnut Ridge Research Building, Morgantown, WV 26505, USA}
%\email{emmanuel.fonseca@nanograv.org}
\author{Gabriel E. Freedman}
\affiliation{NASA Goddard Space Flight Center, Greenbelt, MD 20771, USA}
%\email{gabriel.freedman@nanograv.org}
\author{Nate Garver-Daniels}
\affiliation{Department of Physics and Astronomy, West Virginia University, P.O. Box 6315, Morgantown, WV 26506, USA}
\affiliation{Center for Gravitational Waves and Cosmology, West Virginia University, Chestnut Ridge Research Building, Morgantown, WV 26505, USA}
%\email{nathaniel.garver-daniels@nanograv.org}
\author{Peter A. Gentile}
\affiliation{Department of Physics and Astronomy, West Virginia University, P.O. Box 6315, Morgantown, WV 26506, USA}
\affiliation{Center for Gravitational Waves and Cosmology, West Virginia University, Chestnut Ridge Research Building, Morgantown, WV 26505, USA}
%\email{peter.gentile@nanograv.org}
\author{Joseph Glaser}
\affiliation{Department of Physics and Astronomy, West Virginia University, P.O. Box 6315, Morgantown, WV 26506, USA}
\affiliation{Center for Gravitational Waves and Cosmology, West Virginia University, Chestnut Ridge Research Building, Morgantown, WV 26505, USA}
%\email{joseph.glaser@nanograv.org}
\author{Deborah C. Good}
\affiliation{Department of Physics and Astronomy, University of Montana, 32 Campus Drive, Missoula, MT 59812}
%\email{deborah.good@nanograv.org}
\author{Jeffrey S. Hazboun}
\affiliation{Department of Physics, Oregon State University, Corvallis, OR 97331, USA}
%\email{jeffrey.hazboun@nanograv.org}
\author{Ross J. Jennings}
\altaffiliation{NANOGrav Physics Frontiers Center Postdoctoral Fellow}
\affiliation{Department of Physics and Astronomy, West Virginia University, P.O. Box 6315, Morgantown, WV 26506, USA}
\affiliation{Center for Gravitational Waves and Cosmology, West Virginia University, Chestnut Ridge Research Building, Morgantown, WV 26505, USA}
%\email{ross.jennings@nanograv.org}
\author{Megan L. Jones}
\affiliation{Center for Gravitation, Cosmology and Astrophysics, Department of Physics and Astronomy, University of Wisconsin-Milwaukee,\\ P.O. Box 413, Milwaukee, WI 53201, USA}
%\email{megan.jones@nanograv.org}
\author{David L. Kaplan}
\affiliation{Center for Gravitation, Cosmology and Astrophysics, Department of Physics and Astronomy, University of Wisconsin-Milwaukee,\\ P.O. Box 413, Milwaukee, WI 53201, USA}
%\email{kaplan@uwm.edu}
\author{Matthew Kerr}
\affiliation{Space Science Division, Naval Research Laboratory, Washington, DC 20375-5352, USA}
%\email{matthew.kerr@nanograv.org}
\author{Michael T. Lam}
\affiliation{SETI Institute, 339 N Bernardo Ave Suite 200, Mountain View, CA 94043, USA}
\affiliation{School of Physics and Astronomy, Rochester Institute of Technology, Rochester, NY 14623, USA}
\affiliation{Laboratory for Multiwavelength Astrophysics, Rochester Institute of Technology, Rochester, NY 14623, USA}
%\email{michael.lam@nanograv.org}
\author{Duncan R. Lorimer}
\affiliation{Department of Physics and Astronomy, West Virginia University, P.O. Box 6315, Morgantown, WV 26506, USA}
\affiliation{Center for Gravitational Waves and Cosmology, West Virginia University, Chestnut Ridge Research Building, Morgantown, WV 26505, USA}
%\email{duncan.lorimer@nanograv.org}
\author{Jing Luo}
\altaffiliation{Deceased}
\affiliation{Department of Astronomy \& Astrophysics, University of Toronto, 50 Saint George Street, Toronto, ON M5S 3H4, Canada}
%\email{jing.luo@nanograv.org}
\author{Ryan S. Lynch}
\affiliation{Green Bank Observatory, P.O. Box 2, Green Bank, WV 24944, USA}
%\email{ryan.lynch@nanograv.org}
\author{Alexander McEwen}
\affiliation{Center for Gravitation, Cosmology and Astrophysics, Department of Physics and Astronomy, University of Wisconsin-Milwaukee,\\ P.O. Box 413, Milwaukee, WI 53201, USA}
%\email{alexander.mcewen@nanograv.org}
\author{Maura A. McLaughlin}
\affiliation{Department of Physics and Astronomy, West Virginia University, P.O. Box 6315, Morgantown, WV 26506, USA}
\affiliation{Center for Gravitational Waves and Cosmology, West Virginia University, Chestnut Ridge Research Building, Morgantown, WV 26505, USA}
%\email{maura.mclaughlin@nanograv.org}
\author{Natasha McMann}
\affiliation{Department of Physics and Astronomy, Vanderbilt University, 2301 Vanderbilt Place, Nashville, TN 37235, USA}
%\email{natasha.mcmann@nanograv.org}
\author{Bradley W. Meyers}
\affiliation{Australian SKA Regional Centre (AusSRC), Curtin University, Bentley, WA 6102, Australia}
\affiliation{International Centre for Radio Astronomy Research (ICRAR), Curtin University, Bentley, WA 6102, Australia}
%\email{bradley.meyers@nanograv.org}
\author{Cherry Ng}
\affiliation{Dunlap Institute for Astronomy and Astrophysics, University of Toronto, 50 St. George St., Toronto, ON M5S 3H4, Canada}
%\email{cherry.ng@nanograv.org}
\author{David J. Nice}
\affiliation{Department of Physics, Lafayette College, Easton, PA 18042, USA}
%\email{niced@lafayette.edu}
\author{Timothy T. Pennucci}
\affiliation{Institute of Physics and Astronomy, E\"{o}tv\"{o}s Lor\'{a}nd University, P\'{a}zm\'{a}ny P. s. 1/A, 1117 Budapest, Hungary}
%\email{timothy.pennucci@nanograv.org}
\author{Benetge B. P. Perera}
\affiliation{Arecibo Observatory, HC3 Box 53995, Arecibo, PR 00612, USA}
%\email{benetge.perera@nanograv.org}
\author{Nihan S. Pol}
\affiliation{Department of Physics, Texas Tech University, Box 41051, Lubbock, TX 79409, USA}
%\email{nihan.pol@nanograv.org}
\author{Henri A. Radovan}
\affiliation{Department of Physics, University of Puerto Rico, Mayag\"{u}ez, PR 00681, USA}
%\email{henri.radovan@nanograv.org}
\author{Scott M. Ransom}
\affiliation{National Radio Astronomy Observatory, 520 Edgemont Road, Charlottesville, VA 22903, USA}
%\email{sransom@nrao.edu}
\author{Paul S. Ray}
\affiliation{Space Science Division, Naval Research Laboratory, Washington, DC 20375-5352, USA}
%\email{paul.ray@nanograv.org}
\author{Ann Schmiedekamp}
\affiliation{Department of Physics, Penn State Abington, Abington, PA 19001, USA}
%\email{ann.schmiedekamp@nanograv.org}
\author{Carl Schmiedekamp}
\affiliation{Department of Physics, Penn State Abington, Abington, PA 19001, USA}
%\email{carl.schmiedekamp@nanograv.org}
\author{Brent J. Shapiro-Albert}
\affiliation{Department of Physics and Astronomy, West Virginia University, P.O. Box 6315, Morgantown, WV 26506, USA}
\affiliation{Center for Gravitational Waves and Cosmology, West Virginia University, Chestnut Ridge Research Building, Morgantown, WV 26505, USA}
\affiliation{Giant Army, 915A 17th Ave, Seattle WA 98122}
%\email{brent.shapiro-albert@nanograv.org}
\author{Ingrid H. Stairs}
\affiliation{Department of Physics and Astronomy, University of British Columbia, 6224 Agricultural Road, Vancouver, BC V6T 1Z1, Canada}
%\email{stairs@astro.ubc.ca}
\author{Kevin Stovall}
\affiliation{National Radio Astronomy Observatory, 1003 Lopezville Rd., Socorro, NM 87801, USA}
%\email{kevin.stovall@nanograv.org}
\author{Abhimanyu Susobhanan}
\affiliation{Max-Planck-Institut f{\"u}r Gravitationsphysik (Albert-Einstein-Institut), Callinstra{\ss}e 38, D-30167 Hannover, Germany\\Leibniz Universit{\"a}t Hannover, D-30167 Hannover, Germany}
%\email{abhimanyu.susobhanan@nanograv.org}
\author{Joseph K. Swiggum}
\altaffiliation{NANOGrav Physics Frontiers Center Postdoctoral Fellow}
\affiliation{Department of Physics, Lafayette College, Easton, PA 18042, USA}
%\email{joseph.swiggum@nanograv.org}
\author{Stephen R. Taylor}
\affiliation{Department of Physics and Astronomy, Vanderbilt University, 2301 Vanderbilt Place, Nashville, TN 37235, USA}
%\email{stephen.taylor@nanograv.org}
\author{Michele Vallisneri}
\affiliation{ETH Zurich, Institute for Particle Physics and Astrophysics, Wolfgang-Pauli-Strasse 27, 8093 Zurich, Switzerland}
%\email{michele.vallisneri@nanograv.org}
\author{Rutger van~Haasteren}
\affiliation{Max-Planck-Institut f{\"u}r Gravitationsphysik (Albert-Einstein-Institut), Callinstra{\ss}e 38, D-30167 Hannover, Germany\\Leibniz Universit{\"a}t Hannover, D-30167 Hannover, Germany}
%\email{rutger@vhaasteren.com}
\author{Haley M. Wahl}
\affiliation{Department of Physics and Astronomy, West Virginia University, P.O. Box 6315, Morgantown, WV 26506, USA}
\affiliation{Center for Gravitational Waves and Cosmology, West Virginia University, Chestnut Ridge Research Building, Morgantown, WV 26505, USA}
%\email{haley.wahl@nanograv.org}

\begin{abstract}
Pulsar timing arrays search for stochastic processes such as
gravitational waves by comparing pulse time of arrival data for
millisecond pulsars to expectations from a background with a given
power spectral density (PSD).  To make the analysis computationally
tractable, the Bayesian likelihood is usually computed using an
approximation in which the signal is taken to be a sum of Fourier
modes appropriate to the total time of observation, even though the
true signal is not periodic.  We study the difference between
likelihoods computed with this Fourier approximation method for power
law spectra and those computed exactly (or using more-closely spaced
frequencies as a proxy for the exact result) in the NANOGrav 15-year
dataset.  We find that the true marginal likelihoods for power-law
PSDs are on average about half as large as the likelihoods computed
using the Fourier approximation. This could lead to an error of a
factor of two in model comparison. However, in the important
comparison of uncorrelated vs.\ Hellings-Downs correlated models, a
very similar correction appears in both, so the model comparison is
essentially unaffected. We also compare parameter estimation
results for power law PSDs, finding little difference between the
methods.  We briefly discuss spectra with sharper features, for which
the approximation could be much worse.
\end{abstract}

\maketitle

\section{Introduction}

Pulsar timing array (PTA) collaborations such as the North American
Nanohertz Observatory for Gravitational Waves (NANOGrav) monitor a set
of pulsars to look for small deviations in pulse times of arrival
(TOAs) that may be caused by gravitational waves (GWs).
Hellings-Downs (HD) \cite{HD} correlations between different pulsars
provide a characteristic signature of a stochastic GW background
(GWB).  Recent PTA datasets
\cite{NANOGrav:2023hde,EPTA:2023sfo,Zic:2023gta,Xu:2023wog} gave
strong evidence
\cite{NANOGrav:2023gor,EPTA:2023fyk,Reardon:2023gzh,Xu:2023wog} for
such a GWB.

We often analyze PTA data using Bayesian Markov Chain Monte Carlo
(MCMC) methods, which are based on computing the likelihood of the
observed data (in this case the TOAs) given a particular model, such
as a power-law spectrum of stochastic GWs with HD correlations.  By
these methods we can estimate the values of parameters (such as the
amplitude $A$ and slope $\gamma$ of the power law).  We can also make
a comparison between different models, such as the HD model and a
model without inter-pulsar correlations, called CURN for ``common
uncorrelated red noise.''

In principle, for each model with specific parameters, we can compute
a covariance matrix $C$, where $C_{ij}$ gives the correlation between
the $i$th and $j$th TOA.  However, to calculate a likelihood requires
inverting $C$.  The number of rows and columns of $C$ is the total
number of TOAs, nearly 700,000 in the NANOGrav 15-year dataset.
Inverting such a matrix is computationally intractable, so the
usual technique is to approximate $C$ in low-rank form
\cite{Lentati:2012xb} using a basis of ``Fourier modes'' of
frequencies $n/T$, where $T$ is the total time of observation for the
PTA. However, the accuracy of this approximation and its impact on GW
detection remain open questions.

In this paper, we quantitatively evaluate the errors introduced by the
Fourier approximation. First we compute the exact $C$ for the CURN
model and a small number of parameter sets.  We compare it with the
approximate $C$ computed in the usual way with Fourier modes of
frequency $f_n = n/T$, and also with the same approximation but using
modes with frequencies spaced as\footnote{In expressions such as
$1/10T$, we always mean $1/(10T)$, not $(1/10)T$.} $1/10T$ and
$1/100T$ \cite{vanHaasteren:2014faa}.  We find that these give very
good approximations of the exact $C$.  Then we use $1/10T$ spaced
frequencies as a proxy for the exact calculation to quantify the
effect on Bayes factors and parameter estimation due to using the
usual $1/T$ calculation instead of the exact one in GW detection.

The paper is organized as follows.  Sec.~\ref{sec:calculation} reviews
the mathematical framework for the likelihood calculation and
Sec.~\ref{sec:methods} reviews the Fourier approximation usually used
in PTA GW detection analysis. In Sec.~\ref{sec:frequency spacing}, we
compare the exact and approximate covariance matrices after timing
model projection for various Fourier mode spacings. We find that
matrices using $1/10T$ spaced Fourier modes are very close to the
exact matrices. This motivates us to use a $1/10T$ model as a proxy
for the exact likelihood in Sec.~\ref{sec:likelihoods}, where we
continue the comparison by calculating and comparing the likelihoods
produced by the $1/10T$ model with the original $1/T$ analysis.  In
Sec.~\ref{sec:parameters}, we evaluate the effect of the decreased
frequency spacing on Bayesian MCMC parameter recovery. Finally, in
Sec.~\ref{sec:bayes}, we calculate the Bayes factor between exact and
approximate CURN and HD models to see if the $1/T$ model causes any
systematic inaccuracy in determining the preference for HD, a vital
piece of evidence for a GWB.  In Sec.~\ref{sec:conclusion}, we
summarize our results and discuss implications and caveats.

\section{Calculation of likelihoods of PTA data}
\label{sec:calculation}

The fundamental information used in PTA analysis is a vector
$t$, giving the processed times of arrival of pulses at many
different epochs.  Each TOA is written as a deterministic part
plus a stochastic part,
\be
\label{vect}
    t = \tdet + \delta t\,.
\ee
Here $\tdet$ represents the part of the TOA that results from the
best-fit timing model, which includes the pulsar phase, frequency,
spin-down rate, and usually many other effects.  The remainder,
$\delta t$, is the timing residual that contains the stochastic
effects such as GW signals.  We can further decompose $\delta t$ into
three stochastic processes: white noise, per-pulsar red noise, and
GW. The white noise includes the detector's noise and measurement
uncertainties, and is not related to the physical processes we are
interested in. The red noise includes the pulsar's intrinsic
rotational instabilities and unmodeled propagation delays in the
interstellar medium.  The GW component is the effect of a stochastic
gravitational wave background with a given spectrum. Thus we have
\cite{vanHaasteren:2008yh}
\be
    \delta t = \delta \twhite + \delta \tred + \delta \tGW\,.
\ee
Gravitational waves would lead to HD correlations between different
pulsars, but in many cases to speed up the computations we consider the
pulsars only individually, giving the CURN model.

We have separate (one-sided) power-law power spectral densities $S(f)$
for the red noise process of each pulsar and one for the GWB.  Each
has the form
\be\label{eqn:S}
    S(f)=B f^{-\gamma} \quad\text{with}\quad B = \frac{A^2}{12\pi^2} f_1^{\gamma-3}\,,
\ee
where parameter $A$ is the amplitude of the signal, $\gamma$ is
the spectral index of the power law, and the reference frequency $f_1 = 1/\yr$.

As we do not know the real $A$ and $\gamma$ for each process, during
the signal analysis we build a model for intrinsic noise and GWB by
sampling the $A$ and $\gamma$ through a Bayesian MCMC process. Each
pulsar's red noise process and the GWB are described by
their own $A$ and $\gamma$, resulting in 136 total parameters for 67
pulsars and one common GWB.

We then calculate the likelihood of the observed data for each set of $A$ and
$\gamma$. To do this, we need
covariance matrices $C$ for the stochastic processes in $\delta
t$. In the uncorrelated case, there is a covariance matrix for
each pulsar of the form \cite{vanHaasteren:2008yh}
\be
    C = N +\Cred+\CGW\,.
\ee
where $N$ is the white noise covariance matrix,
$\CGW$ is the expectation value of the outer product of $\delta \tGW$:
\be
    \CGW_{ij}=\langle \delta t^{GW}_i, \delta t^{GW}_j \rangle\,,
\ee
and similarly for $\Cred$.

Since the GWB is treated as a red noise common to all pulsars, $\Cred$
and $\CGW$ are calculated using the same equations but different $A$
and $\gamma$ parameters. For the moment, let $C$ denote either
$\CGW$ or $\Cred$, and $S$ and $\gamma$ be those quantities
appropriate to that $C$.  

Via the Wiener-Khinchin theorem, the covariance matrix can be written as a Fourier transform of the power spectral density $S(f)$ \cite{vanHaasteren:2012hj}, 
\be\label{eqn:C1}
    C_{ij}=\int_0^\infty S(f)\cos(f\tau_{ij}) df\,,
\ee
where $\tau_{ij}=2\pi |t_i-t_j|$.  This integral diverges in the usual
case where $\gamma > 1$.  However, the divergence results from modes
whose periods are much longer than the total time of observation of
the PTA, and these cannot be observed.  In Eq.~(\ref{vect}) we
decompose the observed TOAs into the timing model and the residuals,
but the parameters of the timing model are uncertain
\cite{Edwards:2006zg,vanHaasteren:2012hj}.  Since we don't
know the rotational phase of the pulsar or its precise distance from
us, there is an uncertainty of a constant term in the TOAs.  Similarly
the pulsar rotation rate and redshift give a linear uncertainty, and
the spin-down rate and acceleration relative to us give a quadratic
uncertainty.  Thus we are insensitive to any constant, linear, or
quadratic effect in the TOAs.  Often there are many other sources of
uncertainty.

To quantify these effects we the construct an $\NTOA\times N_M$ matrix
$M$, each of whose columns gives the effect of varying one timing
model parameter.  We project out these uncertainties using a matrix
$G$ whose columns are orthonormal and orthogonal to the columns of $M$
\cite{vanHaasteren:2012hj}.  We will be interested not in $C$ but $C'
= G^T C G$, the covariance matrix with the timing-model uncertainty
projected out.  This projection will cure the divergence in the
Wiener-Khinchin integral.

We put Eq.~(\ref{eqn:S}) into Eq.~(\ref{eqn:C1}), contract with $G$ on
each side inside the integral, and expand the cosine in a power
series \cite{vanHaasteren:2014faa}.  We get
\be\label{eqn:Cp1}
C'_{ij} = B \int_0^\infty
G_{ki} f^{-\gamma}\sum_{n=0}^\infty (-1)^n \frac{(f\tau_{kl})^{2n}}{(2n)!}G_{lj}\,,
\ee
where repeated indices are summed over.  We can think of the sum over
$n$ in Eq.~(\ref{eqn:Cp1}) as a power series in the two variables
$t_k$ and $t_l$.  By construction, $G_{ki}$ annihilates constant,
linear, and quadratic functions of $t_k$.  Similarly $G_{lj}$
annihilates such functions of $t_l$.  Thus terms with $n<3$ do not
contribute, and the lowest power of $f$ that enters into
Eq.~(\ref{eqn:Cp1}) is $f^{6-\gamma}$.  As long as $\gamma < 7$, the
integral will converge.

We can do the integral in Eq.~(\ref{eqn:Cp1}) term by term in
$n$.  The indefinite integral is
\be\label{eqn:indefinite}
G_{ki} f^{1-\gamma}\sum_{n=0}^\infty (-1)^n \frac{(f\tau_{kl})^{2n}}{(2n)!(2n+1-\gamma) } G_{lj} \,,
\ee
and the sum can be done, giving
\be\label{eqn:indefinite2}
G_{ki}\frac{f^{1-\gamma}{}_1F_2 \left((1-\gamma)/2; 1/2, (3-\gamma)/2;
  -(f\tau_{kl})^2/4\right)}{1-\gamma}G_{lj}\,,
\ee
where $_1F_2$ is the generalized hypergeometric function.  In
Eq.~(\ref{eqn:indefinite}), the first surviving term has $f^{7-\gamma}$.
This and all subsequent terms vanish in the $f\to 0$ limit.  Thus the
definite integral in Eq.~(\ref{eqn:Cp1}) can be found by taking the
$f\to\infty$ limit of Eq.~(\ref{eqn:indefinite2}), giving
\be\label{eqn:Cp2}
C'_{ij} = B \Gamma(1-\gamma) \sin \left(\frac{\pi \gamma}{2}\right)
G_{ki} \tau_{ij}^{\gamma -1} G_{lj}
\,. 
\ee

Below, we will compare the exact $C'$ matrix with approximations
generated by low-rank methods that consider only a finite frequency
range, typically ending at $f_H = 30/T$.  Higher frequencies are not
generally accessible to PTAs.  For this purpose we need
the integral of Eq.~(\ref{eqn:Cp1}) from 0 to $f_H$.  Thus instead of
taking the $f\to\infty$ limit, we just put $f_H$ into
Eq.~(\ref{eqn:indefinite2}), 
\be\label{eqn:Cp3}
C'_{ij} = B G_{ki}\frac{f^{1-\gamma}{}_1F_2 \left((1-\gamma)/2; 1/2, (3-\gamma)/2;
  -(f_H\tau_{kl})^2/4\right)}{1-\gamma}G_{lj}\,.
\ee

Returning now to the full covariance matrix including white noise, red
noise, and GWB, we can write
\be
C' = N' + \Cpred + \CpGW\,,
\ee
where $\Cpred$ and $\CpGW$ are computed by the above procedure and $N'
= G^T N G$ is the timing-model-projected white noise matrix.

With the $C'$ matrix, the likelihood of one set of parameters is then
calculated by \cite{vanHaasteren:2012hj}
\be
    L(\delta t)=\frac{1}{\sqrt{(2 \pi)^n \det C'}} \exp
    \left(-\frac{1}{2}\delta t^T G C'^{-1} G^T\delta t\right). 
\ee
This direct calculation of likelihood requires us to invert a large and
dense $C'$ matrix. For the CURN models, inter-pulsar correlations of
the GW signals are not considered, so each pulsar has a $C'$ matrix
with $\NTOA-N_M$ rows and columns. For HD models, the pulsars are
computed together with one $C$ matrix with $\sum(\NTOA-N_M)$ rows and
columns, where the sum is over all pulsars. Thus, the CURN model
is much less expensive to do computation on, and will be the focus in
Sec.~\ref{sec:frequency spacing}.

\section{The Fourier Approximation} 
\label{sec:methods}

To enable computation, we are interested in approximating $\Cred+\CGW$
as some matrix $K$ which is easier to invert. To find a suitable  $K$,
the components of red noise processes (including the common GW
process) in the timing residual are first expressed in a Fourier
expansion \cite{Lentati:2012xb},
\be\label{eqn:FT}
    \delta \tGW\approx Fa\,,
\ee
where $a$ is a  vector of Fourier coefficients, $F$ is a
matrix of the Fourier basis functions,
\be\label{eqn:F}
F = \begin{bmatrix}
\cos{(2\pi f_1 t_1)} & \sin{(2\pi f_1 t_1)} & \dots & \cos{(2\pi
  f_\Nfreq t_1)} & \sin{(2 \pi f_\Nfreq t_1)} \\
\cos{(2\pi f_1 t_2)} & \sin{(2\pi f_1 t_2)} & \dots & \cos{(2\pi
  f_\Nfreq t_2)} & \sin{(2 \pi f_\Nfreq t_2)} \\
\vdots & \vdots &\ddots & \vdots & \vdots \\
\cos{(2\pi f_1 t_\NTOA)} & \sin{(2\pi f_1 t_\NTOA)} & \dots & \cos{(2\pi f_\Nfreq t_\NTOA)} & \sin{(2 \pi f_\Nfreq t_\NTOA)} \\
\end{bmatrix}\,,
\ee
and $t_n$ is the $n^{\text{th}}$ TOA.  The number of Fourier
frequencies, $\Nfreq$, is normally selected to be 30, giving 30 $\sin$
modes and 30 $\cos$ modes.  The Fourier coefficients $a$
are taken to be independently distributed with covariance
\be\label{eqn:aa}
\langle a_m a_n \rangle = \delta_{mn} S(f_n)\,.
\ee
Equations~(\ref{eqn:FT}--\ref{eqn:aa}) correspond to treating the GWB
as though it were emitted only at the discrete frequencies $f_n$.  If
we had a periodic signal, indeed only these frequencies could exist,
but in our case we have a timeslice of a non-periodic
signal.\footnote{Of course it is possible to treat any function on
$t\in [0,T)$ as periodic by gluing it to itself.  However, this
  generally introduces a discontinuity and thus a fictitious power
  that declines slowly at high frequencies.}

Using Eqs.~(\ref{eqn:FT}--\ref{eqn:aa}) we can construct an
approximation $\KGW \approx \CGW$:
\be
    \KGW = \langle Fa, (Fa)^T \rangle = F \phiGW F^T\,,
\ee
with $\phi$ a diagonal matrix with elements $S(f_n)$.
We can apply the same Fourier approximation to the red noise process, and sum the two resulting matrices to get:
\be
    K = \KGW + \Kred = F \phi F^T\,,
\ee
with $\phi = \phiGW  + \phired$.

Fourier approximation of the covariance matrices allows us to calculate $C^{-1} \approx (N+K)^{-1}$, avoiding the inversion of the full $C$ matrix using the Woodbury matrix identity: 
\be
    C^{-1} \approx (N+K)^{-1} =N+ F \phi F = N^{-1} - N^{-1}F (\phi ^{-1} +F^T N^{-1} F)^{-1}F^T N^{-1}\,.
\ee
Because $N$ is block diagonal and $\phi$ is relatively small ($2\Nfreq
\times 2\Nfreq$), the time required to calculate a given likelihood is
drastically reduced. However, using the Fourier approximation raises
questions about its accuracy. For example, the Fourier-sum expansion
always produces a periodic function, but the GW spectrum is not
periodic \cite{vanHaasteren:2014faa}. Therefore, in this paper, we
evaluate the accuracy of using the Fourier approximation in GW
detection.

\section{Increasing the sampling frequency numbers}
\label{sec:frequency spacing}

The most straightforward way to evaluate the Fourier approximation
would be to calculate likelihoods exactly and compare them to the
approximate ones.  However, for correlated models this is
computationally intractable, and even for uncorrelated models it takes
several hours to compute one likelihood.  So we use a Fourier method
with more closely spaced frequencies \cite{vanHaasteren:2014faa} as a
bridge between the exact calculation and the usual approximation.

\subsection{Method} \label{subsec:matrix method}

To begin, we do an MCMC run with 2 million samples of a standard CURN
model with 14 red noise Fourier modes and 30 GWB Fourier modes, as
used in Ref.~\cite{NANOGrav:2023gor}.  We discard (burn) the first
300,000 of these to use only the converged part of the chain. We then
consider 200 parameter sets evenly spaced in the MCMC chain file and
compute the exact covariance matrix for each pulsar.  For each
parameter set, we exclude from this collection any pulsar that has any
$\gamma>6$, for the following reason.  Below, we are going to compare
exact timing-model-projected covariance matrices against those
produced by a basis of Fourier modes with various spacings.  The
reason we would expect these matrices to be similar is that PTA
observations are insensitive to frequencies much less than $1/T$,
because of the timing model.  But this idea depends on $\gamma$.  With
$\gamma=7$ or larger, the computation of even the
timing-model-projected covariance matrix diverges as $f\to0$.  When
$\gamma$ is close to 7, the computation converges, but very low
frequencies still contribute significantly. Thus we cannot expect a
similar result for the largest $\gamma$ and we must apply a cutoff.

Furthermore, the only reason these large $\gamma$ (which occur only
for the intrinsic pulsar red noise, not the GWB) exist in our dataset is that it was prepared with a lower
frequency limit of $1/T$.  If lower frequencies had been included, the
large $\gamma$ would have been disfavored because they produce very
large signals at lower frequencies, and these are only somewhat
suppressed by the timing model.

We also also exclude any pulsar with $\gamma<1$.  Such a noise
spectrum is unphysical because it would lead to an infinite amount of
gravitational wave power at high frequencies.

We compute the exact timing-model projected covariance matrices $C'$
for the remaining parameter sets, according to Eq.~\ref{eqn:Cp1}, for
both intrinsic red noise and GW using the parameters ($\gamma$ and
$A$) for the individual pulsars and the GWB.\footnote{While the chain
from which the samples were taken used only 14 modes in the GWB, when 
we computed the exact covariance matrices we set the upper cutoff to
$30/T$, and when we computed the approximate covariance matrices we
used 30 frequencies.  This has little effect on the distance between
matrices that we study here.  In the likelihood and Bayes factor
computations below we always used 14 GWB modes.}\footnote{For each pulsar,
this calculation requires evaluating $_1F_2$ in
Eq.~(\ref{eqn:indefinite2}) about $\NTOA^2$ times, which is
computationally expensive. Therefore, for each $\gamma$, we
precalculate $10^6$ values of this function for $f\tau$ evenly spaced
between 0 and 60$\pi$ and use linear interpolation.}  Then we
calculate the corresponding approximate $C$ matrices in the usual way
with frequencies spaced by $1/T$, and also with frequencies spaced by
$1/10T$ and $1/100T$.  We call these $K_1$, $K_{10}$ and $K_{100}$
respectively.  In both $1/10T$ and $1/100T$ cases, the frequency
starts at $1/10T$. In all cases, the highest frequency is $30/T$. We
project out the timing model uncertainty in the $K$ by calculating
$G^TKG$ in each case. Then, we calculate the distance between the
exact $C'$ and each approximation using the Frobenius norm (the
square root of the sum of the squares of the differences of the
elements).

\subsection{Results} \label{subsec:matrix results}
\begin{figure}
    \centering
    \includegraphics[width=0.5\textwidth]{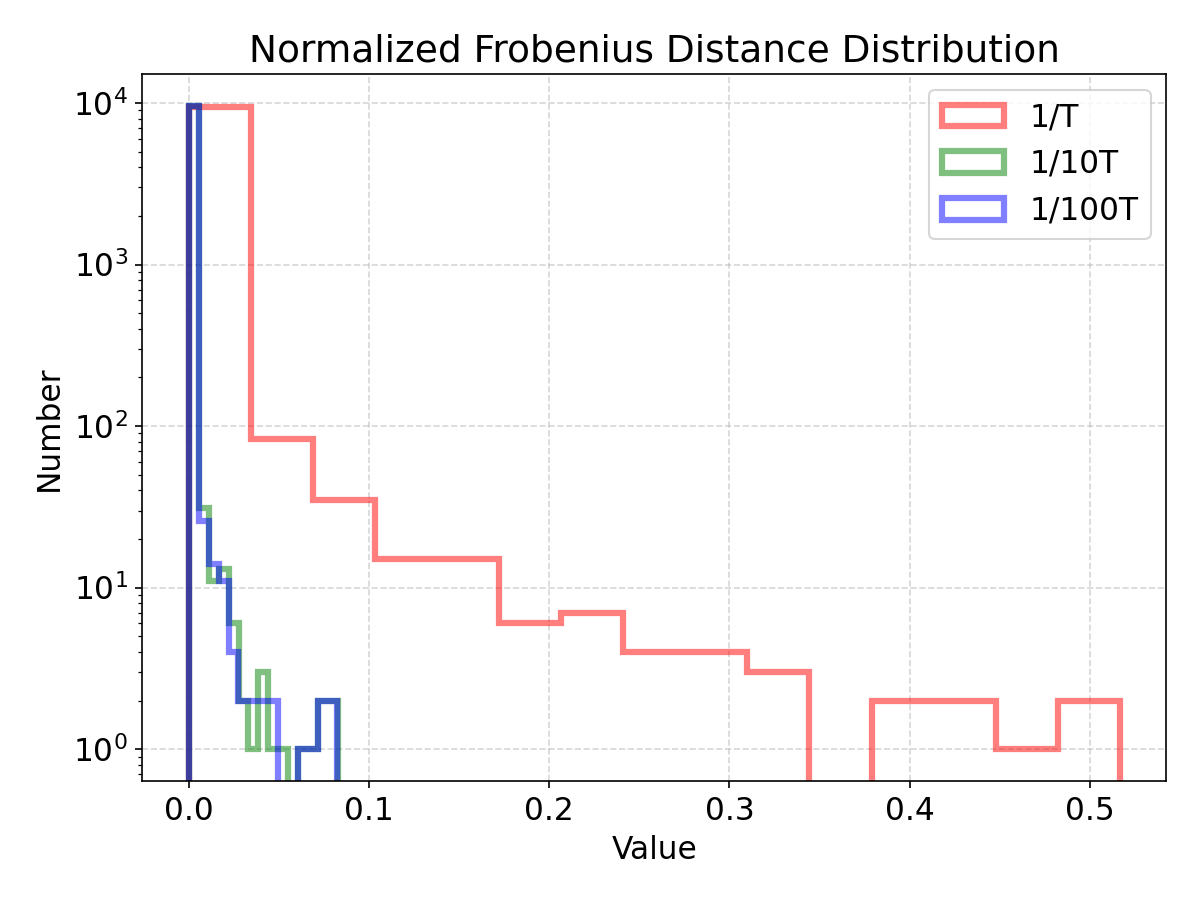}
    \caption{Histogram of Frobenius norms of the difference between
      the exact and 1/T, 1/10T, and 1/100T spaced Fourier matrices,
      normalized by dividing by the norm of the exact matrix.
      There is one matrix for each of the 67 pulsars and each of the
      200 samples from a CURN chain, but we have excluded cases with
      red noise $\gamma$ less than 1 or greater than 6. The total
      number of matrices calculated is 9596. The 1/T matrices give a
      poor approximation.  The other two are much better and similar
      to each other. }
    \label{fig:Histogram}
\end{figure}

Adding more sampled frequencies should, in principle, make the Fourier
matrices converge to the exact matrices. The distribution of the
Frobenius distances of $C'$ is shown in Figure~\ref{fig:Histogram}.
The mean and standard deviation of these norms are shown in Table~\ref{tab:fro}.
\begin{table}
\centering
\begin{tabular}{|l|l|l|l|}
\hline
     & $K_{1}$ & $K_{10}$ & $K_{100}$ \\ \hline
Mean & 0.00333   & 0.000240    & 0.000214    \\ \hline
std  & 0.0204   & 0.00250    & 0.00242     \\ \hline
\end{tabular}
\caption{Mean and standard deviation of the normalized Frobenius norms
  of matrix difference between exact $C'$ matrices and 1/T, 1/10T and
  1/100T Fourier approximations.}
\label{tab:fro}
\end{table}
The Frobenius distance decreases significantly from $K_1$ to $K_{10}$,
and does not change significantly between $K_{10}$ and $K_{100}$.
Thus we see that the even finer spacing does not improve the
approximation.  The reason for the residual difference is that even
our $K_{100}$ has a low-frequency cut off at $1/10T$.  When $\gamma$ is
close to our maximum value of 6, the PTA is still somewhat sensitive
to even lower frequencies, which are included in the exact calculation
but not in any of our approximations.

Since the $K_{10}$ matrices are quite good approximations, we will use
the $1/10T$ approximation as a proxy for the exact calculation and
compare it to the usual $1/T$ result.

\section{Likelihood Differences}
\label{sec:likelihoods}
\begin{figure}
    \centering
    \includegraphics[width=0.5\textwidth]{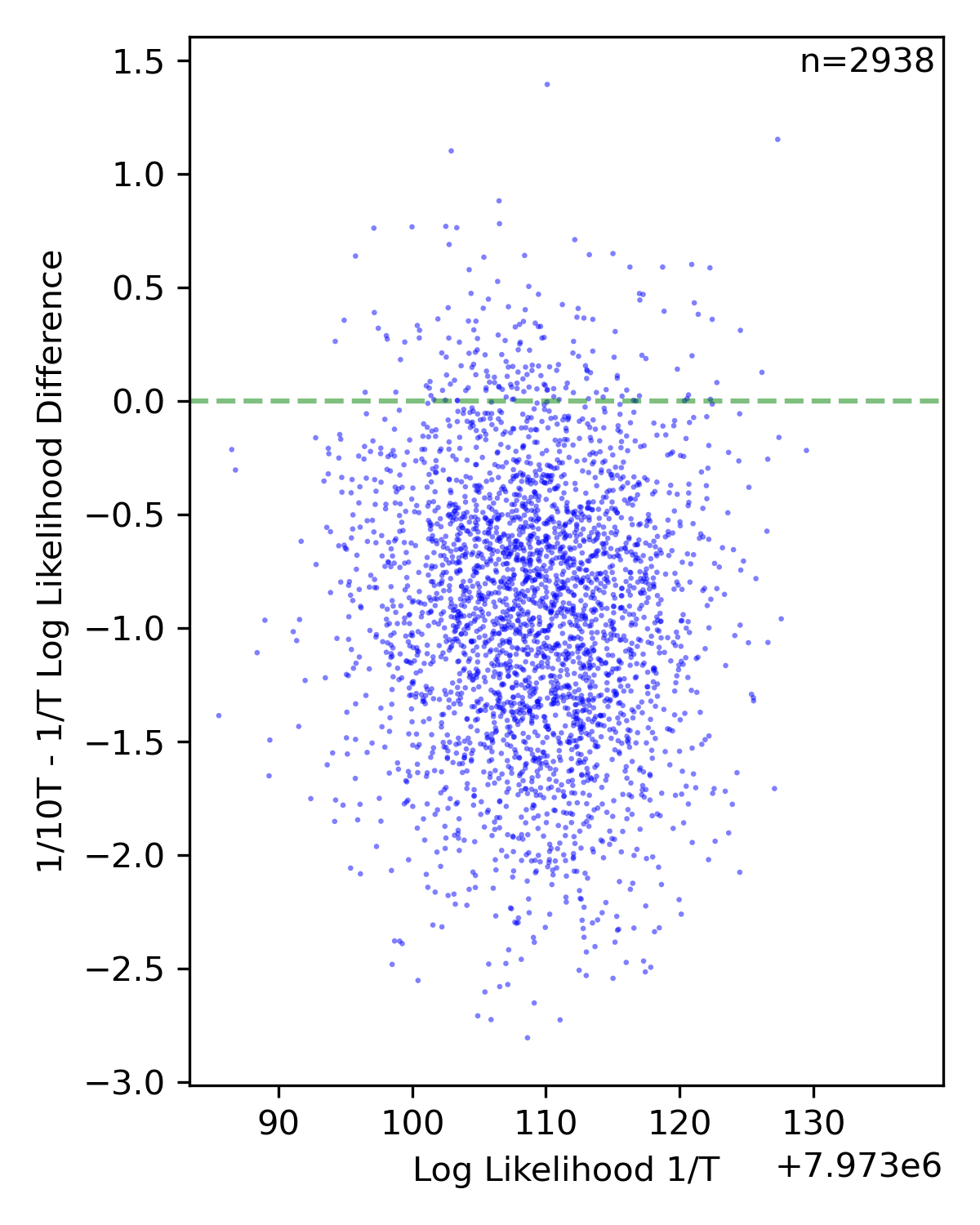}
    \caption{ The difference between the $1/10T$ and the $1/T$ CURN
      likelihoods for the entire PTA plotted against the $1/T$
      likelihood for 2938 samples.  The green dashed line
      shows where they are equal.}
    \label{fig:lndiffCURN}
\end{figure}

\begin{figure}
    \centering
    \includegraphics[width=0.5\textwidth]{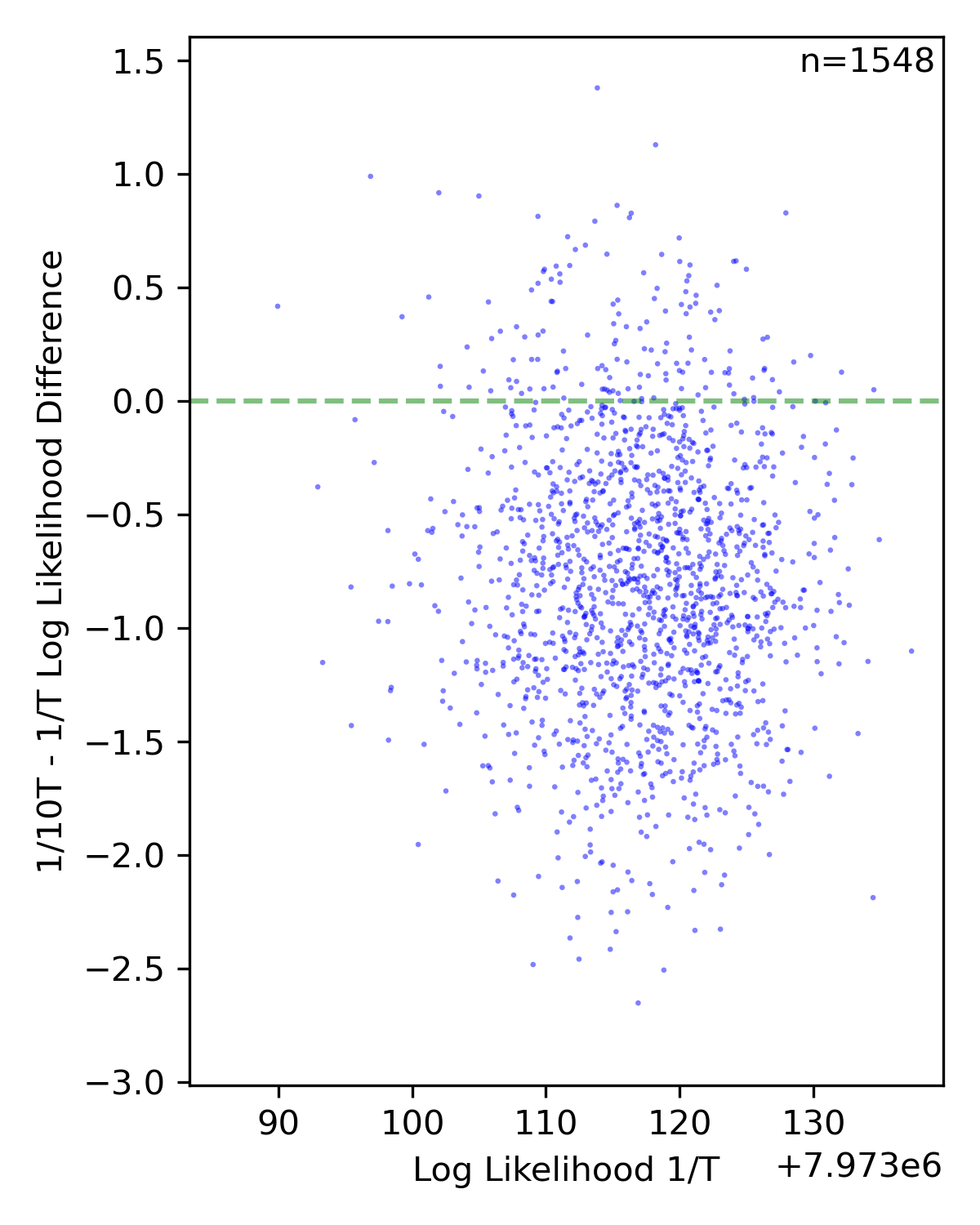}
    \caption{As in Fig.~\ref{fig:lndiffCURN} for HD likelihoods with
      1548 samples.}
    \label{fig:lndiffHD}
\end{figure}

To determine the importance of the error introduced by the use of the
Fourier modes, we focus on the end result of the analysis: the
likelihood of the model with given parameters.  Now we use the $1/10T$
model as our standard, which enables us to analyze more samples than
just the 200 we used earlier in the exact computation.  Following the
reasoning in Section \ref{subsec:matrix method}, we create new chains
with the red noise and gravitational wave power spectral indices
limited to between 0 and 6, rather than the default of 0 and
7.  This chains are otherwise the same as the chain described in
Section \ref{subsec:matrix method}.  We make a CURN chain with 5
million samples and an HD chain with 2.3 million.  After burning the
first 300,000 samples in each chain and thinning by the autocorrelation
distance, we find 2938 independent CURN samples and 1548 independent
HD samples.  We then compare each of these likelihoods with the
likelihood for the same parameters computed with the $1/10T$ model.
These differences are shown in Figures \ref{fig:lndiffCURN} and
\ref{fig:lndiffHD}.

In both cases the $1/10T$ natural log likelihood ranges from around 1
higher to around 3 lower than the $1/T$ likelihood.  In the great
majority of cases the $1/10T$ log likelihood is lower.  For CURN, the
average difference is -0.928 with standard deviation 0.598.  and for
HD the average difference is -0.785 with standard deviation 0.604.
Thus it appears the standard method overestimates the log likelihood
by roughly $0.9\pm 0.6$ unit for CURN and $0.8\pm 0.6$ unit for HD.
There does not appear to be any correlation between the originally
sampled likelihood and the difference in likelihoods.

\section{Parameter Estimation} \label{sec:parameters}
An important use for likelihood computation is to give a probability
distribution of model parameters, here the $A$ and $\gamma$ of the
GWB.  Assuming the above model (GWB, intrinsic pulsar red noise, and
white noise) is the correct explanation for the observed data, this
distribution tells us the posterior probability for each $(A, \gamma)$
according to the data.  Inaccuracies in the likelihood calculation
might lead to incorrect estimation of parameters.  The degree of
likelihood differences found in the previous section are large enough
to lead to important corrections if they are correlated with $A$ and
$\gamma$.  In this section we see whether there is such an effect.

\begin{figure}
    \centering
    \includegraphics[width=0.5\textwidth]{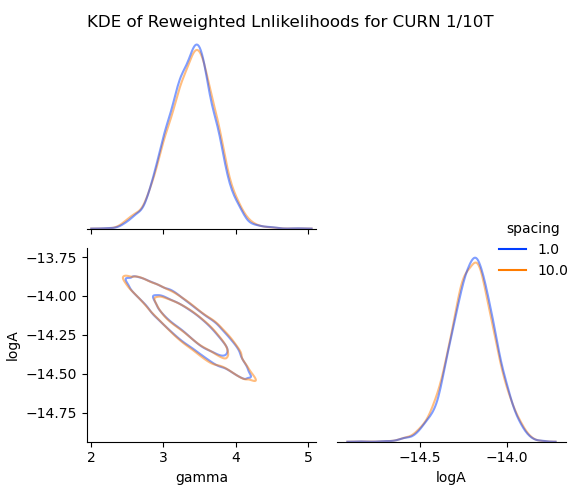}
    \caption{Comparison of parameter estimations using $1/T$ and
      $1/10T$ spacing in the CURN model. The originally sampled
      parameter space is shown in blue and the $1/10T$ reweighted
      parameter space in shown in orange. A kernel density estimator
      is used to smooth out the distribution. The contours show $1
      \sigma$ and $2 \sigma$ levels.}
    \label{fig:kdeCURN}
\end{figure}

\begin{figure}
    \centering
    \includegraphics[width=0.5\textwidth]{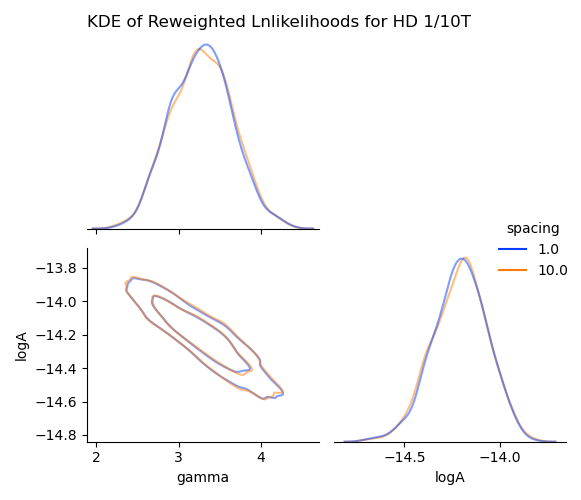}
    \caption{As in Fig.~\ref{fig:kdeCURN} but for the HD model}
    \label{fig:kdeHD}
\end{figure}

Rather than redo our Bayesian MCMC sampling for the $1/10T$ case, we
use the chain described in Sec.~\ref{sec:likelihoods} and do the
parameter estimation by reweighting \cite{Payne_2019,Hourihane_2023}.
For each parameter set $\theta$, consisting of the $A$ and $\gamma$
for the GWB and all the intrinsic pulsar red noises, we compute a
weight
\be\label{eqn:w}
w(\delta t,\theta)=L_{10}(\delta t|\theta)/L_{1}(\delta t|\theta)\,,
\ee
where $L_1$ and $L_{10}$ denote the likelihoods computed
according to the $1/T$ and $1/10T$ models respectively.  (We use the
same prior for both cases.)  This gives a new chain, whose samples now
have weights indicating how likely they are to arise according to the
$1/10T$ chain.  From this chain we compute the posterior distribution
of parameters $A$ and $\gamma$, smoothed by kernel density
estimation. Figure \ref{fig:kdeCURN}
shows the result for the CURN model, and Figure \ref{fig:kdeHD}
shows the result for HD.

In both cases there is little difference between the parameter
distributions computed with $1/T$ and $1/10$ spacing.  Thus the
Fourier approximation is not significantly impacting the sampling of
the GWB parameter space.

\section{Bayes factor} \label{sec:bayes}

Now we investigate the impact of the Fourier approximation on the
Bayes factors used for model selection. If the likelihoods for a given
model are consistently misestimated in the approximation, it will
affect the marginal likelihood of the model, and thus the Bayes
factor when compared with other models. To understand the effect on
the marginal likelihood of the approximation, we find the Bayes factor
between otherwise identical models that differ only in the spacing of
their Fourier modes. We perform this analysis independently for the
common uncorrelated red noise model and the Hellings-Downs model using
the same reweighting method as in the previous section.  Computation
of Bayes factors using reweighting can be considered one step of the
``steppingstone'' \cite{Xie:SS} method.

The marginal likelihood can be written
\be
\cZ = \int d\theta L(\delta|\theta) \pi(\theta)\,,
\ee
where $\theta$ denotes the parameters $A$ and $\gamma$ and
$\pi$ is the prior.   The normalized posterior is
\be
p(\theta) = \frac{L(\delta|\theta) \pi(\theta)}{\cZ}\,.
\ee
Then the marginal likelihood in the $1/10T$ model can be written
\cite{Hourihane_2023}
\be\label{eqn:Z10}
\cZ_{10} = \cZ_1 \int d\theta w(\delta t,\theta) p_1(\theta)\,,
\ee
with $w$ defined in Eq.~(\ref{eqn:w}).  We can approximate
Eq.~(\ref{eqn:Z10}) using our $1/T$ chain, to find the Bayes factor to
for the $1/10T$ calculation vs.\ the $1/T$ calculation,
\be\label{eqn:B}
    \mathcal{B}^{10}_1 = \frac{\cZ_{10}}{\cZ_1} \approx \overline{w}\,,
\ee
where $\overline{w}$ is the average of $w(\delta t,\theta)$ over the
chain.

We calculated the Bayes factor for the $1/10T$ case over the $1/T$
case using Eq.~(\ref{eqn:B}) for the CURN and HD models.  For CURN, we
found $0.472$ and for HD found $0.549$ in general agreement with
Figs.~\ref{fig:lndiffCURN} and \ref{fig:lndiffHD}.  Considering the
$1/10T$ result as a proxy for the exact calculation, we conclude that
the $1/T$ spacing is consistently overestimating the likelihood of the
model by roughly a factor of 2. However, because the overestimation is
very similar for both CURN and HD models, there is no significant
impact on the Bayes factor between CURN and HD as given in the
NANOGrav 15yr GWB paper \cite{NANOGrav:2023gor}.

\section{Conclusion}
\label{sec:conclusion}

We evaluated the Fourier approximation used to detect low-frequency
gravitational waves from PTAs. We started by calculating the exact
timing-model projected covariance matrices for a group of sampled
parameters from a model using common uncorrelated red noise, and
showed that $1/10T$-spaced Fourier modes provides a close
approximation to the exact model. We then calculated the likelihood
differences between $1/10T$ and $1/T$ models. While some log
likelihoods computed using $1/10T$ were larger, most were smaller,
typically by 1 or 2 units.

While this likelihood difference could in principle be important, in
fact it did not make any significant change to the parameter
estimation results.  We also compared the Bayes factor of these two
models and showed that the $1/T$ model is consistently overestimating
the marginal likelihood by about a factor of 2. But since this
overestimation is consistent across CURN and HD models, the Bayes
factor between CURN and HD is not significantly affected.

There are a few caveats.  Here we studied only power law spectra.  The
$1/T$ analysis may be much worse if used with spectra with more
features.  In particular a spectrum supported only in a narrow range
of frequencies lying inside the $1/T$ interval between Fourier mode
frequencies would not be seen at all in the standard analysis.

We also kept an upper cutoff of $30/T$ for all analyses.  This does
not have much effect on the GWB, which falls as $f^{-\gamma}$ with
$\gamma \ge 2$  as seen in the parameter recovery plots.  However,
intrinsic pulsar red noise spectra sometimes have much smaller
$\gamma$ and thus extend to high frequencies.  In fact if one were to
take seriously the pulsars with $\gamma<1$, the covariance matrix
would diverge at high frequency.

\section*{Data Availability}

The MCMC chains used in this paper were computed using
\textsc{enterprise}, available at
\url{https://github.com/nanograv/enterprise}, and
\textsc{ptmcmcsampler}, available at
\url{github.com/nanograv/PTMCMCSampler}.
The MCMC chains, the data used in the figures, and the software
developed specifically for this paper will be available at Zenodo.
% are available at \cite{dataatzenodo}.

\section*{Acknowledgments}

P.R.B.\ is supported by the Science and Technology Facilities Council, grant number ST/W000946/1.
Pulsar research at UBC is supported by an NSERC Discovery Grant and by CIFAR.
K.C.\ is supported by a UBC Four Year Fellowship (6456).
T.D.\ and M.T.L.\ received support by an NSF Astronomy and Astrophysics Grant (AAG) award number 2009468 during this work.
E.C.F.\ is supported by NASA under award number 80GSFC24M0006.
D.C.G.\ is supported by NSF Astronomy and Astrophysics Grant (AAG) award \#2406919.
D.R.L.\ and M.A.M.\ are supported by NSF \#1458952.
M.A.M.\ is supported by NSF \#2009425.
The Dunlap Institute is funded by an endowment established by the David Dunlap family and the University of Toronto.
K.D.O.\ was supported in part by NSF Grant No.\ 2207267.
T.T.P.\ acknowledges support from the Extragalactic Astrophysics Research Group at E\"{o}tv\"{o}s Lor\'{a}nd University, funded by the E\"{o}tv\"{o}s Lor\'{a}nd Research Network (ELKH), which was used during the development of this research.
H.A.R.\ is supported by NSF Partnerships for Research and Education in Physics (PREP) award No.\ 2216793.
S.M.R.\ and I.H.S.\ are CIFAR Fellows.
Portions of this work performed at NRL were supported by ONR 6.1 basic research funding.
S.R.T.\ acknowledges support from NSF AST-2007993.
S.R.T.\ acknowledges support from an NSF CAREER award \#2146016.

The NANOGrav collaboration was supported
by the National Science Foundation Physics Frontiers Center award
No.\ 2020265.  The authors acknowledge the Tufts University High
Performance Compute
Cluster (\url{https://it.tufts.edu/high-performance-computing}) which was
utilized for the research reported in this paper.

\bibliography{references}{}

\end{document}